\documentstyle[epsf]{cgf}

\volume{18}
\pnumber{1}
\pagerange{17--26}
\renewcommand{\year}{1999}

\newcommand{\figureSimple}[1]{
  \begin{figure}[t]{\sloppy \footnotesize#1}\end{figure}
}

\newcommand{\figureTwoCol}[1]{
  \begin{figure*}[tb]{\sloppy \footnotesize#1}\end{figure*}
}

\newcommand{\tableSimple}[1]{
  \begin{table}[tb]{\sloppy \footnotesize#1}\end{table}
}

\title{Animating Sand, Mud, and Snow}

\author[R. W. Sumner, J. F. O'Brien, and J. K. Hodgins]
{Robert W. Sumner \hfill James F. O'Brien \hfill Jessica K. Hodgins \\
College of Computing and Graphics, Visualization, and Usability Center \\
Georgia Institute of Technology 
}

\begin{document}

\maketitle

\begin{abstract}
  Computer animations often lack the subtle environmental changes that
  should occur due to the actions of the characters.  Squealing car
  tires usually leave no skid marks, airplanes rarely leave jet trails
  in the sky, and most runners leave no footprints.  In this paper, we
  describe a simulation model of ground surfaces that can be deformed
  by the impact of rigid body models of animated characters.  To
  demonstrate the algorithms, we show footprints made by a runner in
  sand, mud, and snow as well as bicycle tire tracks, a bicycle crash,
  and a falling runner.  The shapes of the footprints in the three
  surfaces are quite different, but the effects were controlled
  through only five essentially independent parameters.  To assess the
  realism of the resulting motion, we compare the simulated footprints
  to human footprints in sand.
  
\end{abstract}




\section{Introduction}

To become a communication medium on a par with movies, computer
animations must present a rich view into an artificial world.  Texture
maps applied to three-dimensional models of scenery help to create
some of the required visual complexity.  But static scenery is only
part of the answer; subtle motion of many elements of the scene is
also required.  Trees and bushes should move in response to the wind
created by a passing car, a runner should crush the grass underfoot,
and clouds should drift across the sky.  While simple scenery and
sparse motion can sometimes be used effectively to focus the attention
of the viewer, missing or inconsistent action may also distract the
viewer from the plot or intended message of the animation.  One of the
principles of animation is that the viewer should never be
unintentionally surprised by the motion or lack of it in a
scene\cite{Thomas:1984:DA}.
  
\figureSimple{
  \centerline{\epsfxsize=\columnwidth \epsfbox{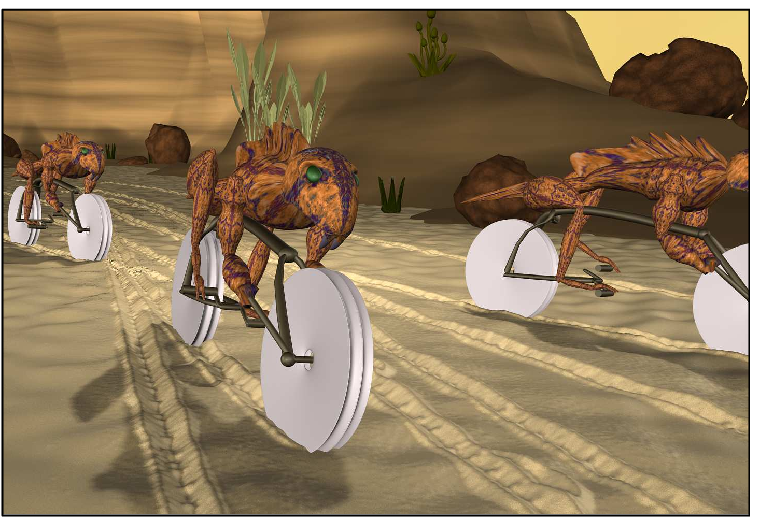}}
  \caption{
    Image of tracks left in the sand by a group of fast moving, 
    alien bikers.  
  }\label{fig:AlienBikes}
}

Subtle changes in the scenery may also convey important information
about the plot or scene context to the viewer.  For example,
figure~\ref{fig:AlienBikes} shows an image of alien bikers riding
across a desert landscape.  The presence of tracks makes it clear that
the ground is soft sand rather than hard rock, and that other bikers
have already passed through the area.  Figure~\ref{fig:NoSand} shows
the same scene without the sand.  In addition to being visually less
interesting, the altered image lacks some of the visual cues that help
the viewer understand the scene.

\figureSimple{
  \centerline{\epsfxsize=\columnwidth \epsfbox{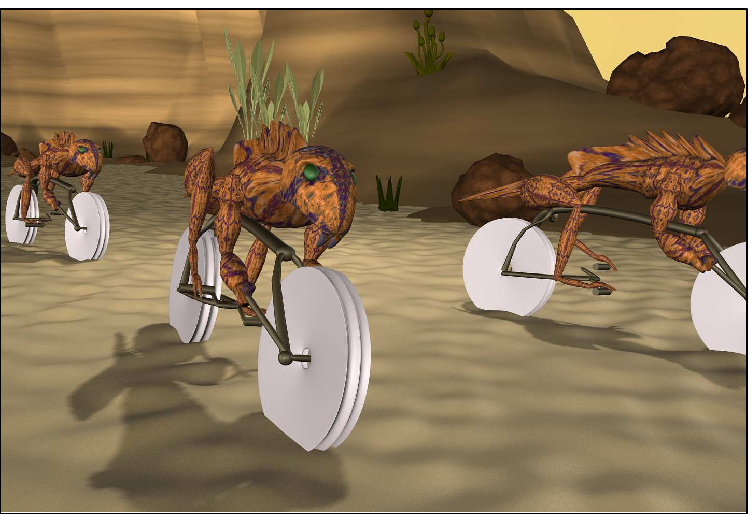}}
  \caption{
    Same image as shown in figure~\ref{fig:AlienBikes} but without the
    simulated tracks in the sand.   
  }\label{fig:NoSand}
}

Movie directors face a related problem because they must ensure that
the viewer is presented with a consistent view of the world and the
characters.  An actor's clothing should not inexplicably change from
scene to scene, lighting should be consistent across edits, and such
absent, unexpected, or anachronistic elements as missing tire tracks,
extra footprints, or jet trails must be avoided.  The risk of
distracting the viewer is so great that one member of the director's
team, known as a ``continuity girl,'' ``floor secretary,'' or ``second
assistant director,'' is responsible solely for maintaining
consistency\cite{Reisz:1994:TFE}.

Maintaining consistency is both easier and harder in computer
animation.  Because we are creating an artificial world, we can
control the lighting conditions, layout, and other scene parameters
and recreate them if we need to ``shoot'' a fill-in scene later.
Because the world is artificial, however, we may be tempted to
re\-arrange objects between scenes for best effect, thereby creating a
series of scenes that could not coexist in a consistent world.
Computer-generated animations and special effects add another facet to
the consistency problem because making models that move and deform
appropriately is a lot of work.  For example, most animated figures do
not leave tracks in the environment as a human actor would and special
effects artists have had to work hard to create such subtle but
essential effects as environment maps of flickering flames.  Because
each detail of the scene represents additional work, computer graphics
environments are often conspicuously clean and sparse.  The approach
presented here is a partial solution to this problem; we create a more
interesting environment by allowing the character's actions to change
a part of the environment.

In this paper, we describe a model of ground surfaces and explain how
these surfaces can be deformed by characters in an animation.  The
ground ma\-terial is modeled as a height field formed by vertical
columns.  After the impact of a rigid body model, the ground material
is deformed by allowing compression of the material and movement of
material between the columns.  To demonstrate the algorithms, we show
the creation of footprints in sand, mud, and snow.  These surfaces are
created by modifying only five essentially independent parameters of
the simulation.  We evaluate the results of the animation through
comparison with video footage of human runners and through more
dramatic patterns created by bicycle tire tracks
(figure~\ref{fig:AlienBikes}), a falling bicycle
(figure~\ref{fig:Bike}), and a tripping runner
(figure~\ref{fig:Trip}).

\section{Background}

Several researchers have investigated the use of pro\-cedural
techniques for generating and animating background elements in
computer-generated scenes.  Although we are primarily interested in
techniques that allow the state of the environment to be altered in
response to the motions of an actor, methods for animating or modeling
a part of the environment independent of the movements of the actors
are also relevant because they can be modified to simulate
interactions.

The first example of animated ground tracks for computer animation was
work done by Lundin\cite{Lundin:1984:MS,SVR:100}. He describes how
footprints can be created efficiently by rendering the underside of
an object to create a bump map and then applying the bump map to the
ground surface to create impressions where the objects have contacted
the ground.

The work most closely related to ours is that of Li and
Moshell\cite{Li:1993:MSR}.  They developed a model of soil that allows
interactions between the soil and the blades of digging machinery.
Soil spread over a terrain is modeled using a height field, and soil
that is pushed in front of a bulldozer's blade is modeled as discrete
chunks.  Although they discount several factors that contribute to
soil behavior in favor of a more tractable model, their technique is
physically based and they arrive at their simulation formulation after
a detailed analysis of soil dynamics.  As the authors note, actual
soil dynamics is complex and their model, therefore, focuses on a
specific set of actions that can be performed on the soil, namely the
effect of horizontal forces acting on the soil causing displacements
and soil slippage.  The method we present here has obvious
similarities to that of Li and Moshell, but we focus on modeling a
different set of phenomena at different scales.  We also adopt a more
appearance-based approach in the interest of developing a technique
that can be used to model a wide variety of ground materials for
animation purposes.

Another method for modeling the appearance of ground surfaces is
described by Chanclou, Luciani, and Habibi\cite{Chanclou:1996:PML}.
They use a simulation-based ground surface model that behaves
essentially like an elastic sheet.  The sheet deforms plasticly when
acted on by other objects.  While their model allows objects to make
smooth impressions in the ground surface, they do not describe how
their technique could be used to realistically model real world ground
materials.

Nishita and his colleagues explored modeling and rendering of snow
using metaballs\cite{Nishita:1997:MRM}.  Their approach allowed them
to model snow on top of objects and
drifts to the side of objects.  They also developed a method
for realistically rendering snow that captured effects due to multiple
levels of light scattering.

Other environmental effects that have been animated include water,
clouds and gases\cite{Ebert:1994:TMP,Stam:1995:DFO,Foster:1997:MMT},
fire\cite{Stam:1995:DFO,Chiba:1994:TDV},
lightning\cite{Reed:1994:VSL}, and leaves blowing in the
wind\cite{Wejchert:1991:AA}.  Among these, water has received the most
attention.  Early work by Peachey\cite{Peachey:1986:MWS} and by
Fournier and Reeves\cite{Fournier:1986:SMO} used procedural models
based on specially designed wave functions to model ocean waves as
they travel and break on a beach.  Later work by Kass and
Miller\cite{Kass:1990:RSF} developed a more general approach using
shallow water equations to model the behavior of water under a wider
variety of conditions.  Their model also modified the appearance of a
sand texture as it became wet.  O'Brien and
Hodgins\cite{OBrien:1995:DSS} extended the work of Kass and Miller to
allow the behavior of the water simulation to be affected by the
motion of other objects in the environment and to allow the water to
affect the motion of the other objects.  They included examples of
objects floating on the surface and simulated humans diving into pools
of water.  More recently Foster and Metaxas\cite{Foster:1996:RAL} used
a variation of the three-dimensional Navier-Stokes equations to model
fluids.  In addition to these surface and volumetric approaches,
particle-based methods have been used to model water spray and other
loosely packed materials.  Supplementing particle models with
inter-particle dynamics allows a wider range of phenomena to be
modeled.  Examples of these systems include
Reeves\cite{Reeves:1983:PST}, Sims\cite{Sims:1990:PAR}, Miller and
Pearce\cite{Miller:1989:GDC}, and Terzopoulos, Platt, and
Fleischer\cite{Terzopoulos:1989:HMD}.

Simulation of interactions with the environment can also be used to
generate still models.  Several researchers have described techniques
for generating complex plant models from grammars describing how the
plant should develop or grow over time.  M{\v{e}}ch and
Prusinkiewicz\cite{Mech:1996:VMP} developed techniques for allowing
developing plants to affect and be affected by their environment.
Dorsey and her colleagues\cite{Dorsey:1996:MRM,Dorsey:1996:FCA} used
simulation to model how an object's surface changes over time as
environmental factors act on it.
 

\section {Simulation of Sand, Mud, and Snow}

In this paper, we present a general model of a deformable ground
material.  The model consists of a height field defined by vertical
columns of material.  Using displacement and compression algorithms,
we animate the deformations that are created when rigid geometric
objects impact the ground material and create footprints, tire tracks,
or other patterns on the ground.  The properties of the model can be
varied to produce the behavior of different ground materials such as
sand, mud, and snow.

\subsection {Model of Ground Material}

\figureSimple{
  \centerline{\epsfxsize= 0.8 \columnwidth \epsfbox{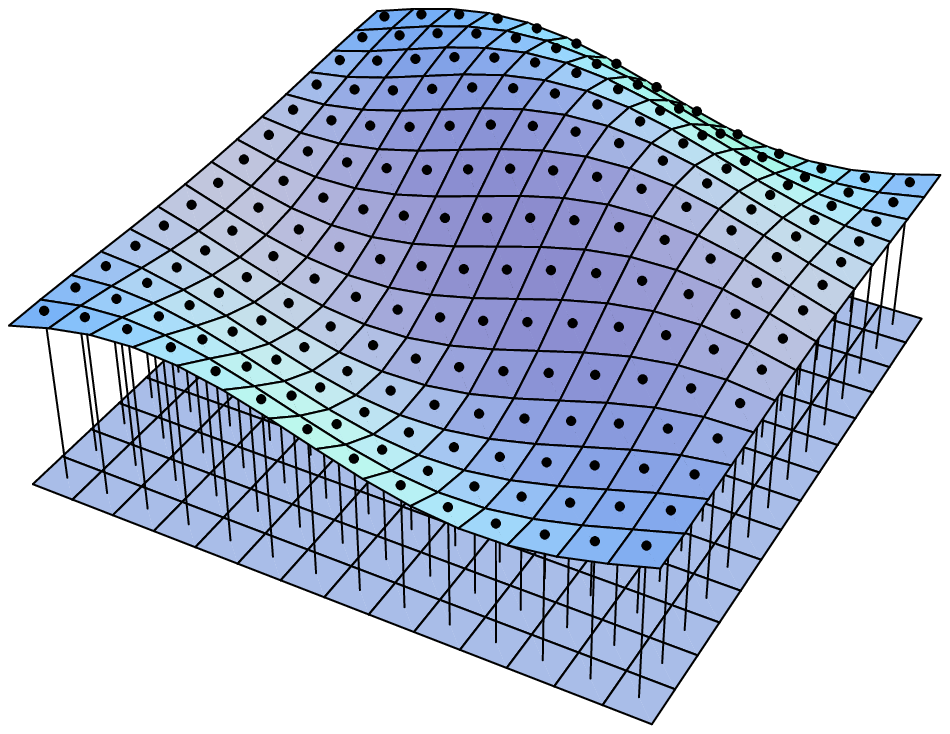}}
  \caption{
    The uniform grid forms a height field that defines the ground
    surface.  Each grid point within the height field represents a
    vertical column of ground material with the top of the column
    centered at the grid point.
  }\label{fig:HeightField}
}

\figureTwoCol{
  \centerline{\epsfxsize=\textwidth \epsfbox{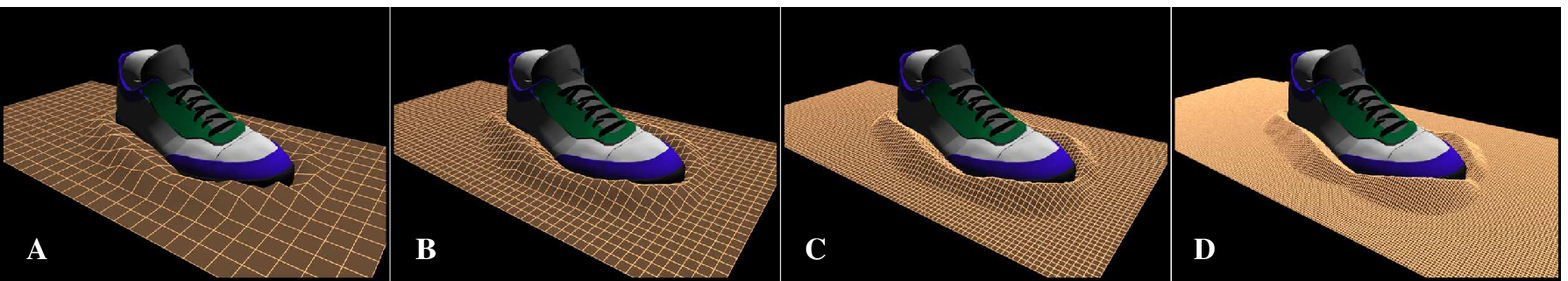}}
  \caption{
    Footprint in sand computed with different grid resolutions.  {\em
    (A)} 20\,mm, {\em (B)} 10\,mm, {\em (C)} 5\,mm, and {\em (D)}~2.5\,mm. As
    the grid resolution increases, the shape of the footprint is
    defined more clearly but its overall shape remains the same.
  }\label{fig:Resolution}
}

Our simulation model discretizes a continuous volume of ground
material by dividing the surface of the volume into a uniform
rectilinear grid that defines a height field
(figure~\ref{fig:HeightField}).  The resolution of the grid must be
chosen appropriately for the size of the desired features in the
ground surface.  For example, in figure~\ref{fig:AlienBikes} the
resolution of the grid is 1\,cm and the bicycles are approximately
2\,meters long with tires~8\,cm wide.  Though the resolution of the
grid determines the size of the smallest feature that can be
represented, it does not otherwise dramatically affect the shape of
the resulting terrain (figure~\ref{fig:Resolution}).

Initial conditions for the height of each grid point can be created
procedurally or imported from a variety of sources.  We implemented
initial conditions with noise generated on an integer lattice and
interpolated with cubic Catmull-Rom splines (a variation of a
two-dimensional Perlin noise function described by
Ebert\cite{Ebert:1994:TMP}).  Terrain data or the output from a
modeling program could also be used for the initial height field.
Alternatively, the initial conditions could be the output of a
previous simulation run.  For example, the trampled surface of a
public beach at the end of a busy summer day could be modeled by
simulating many crisscrossing paths of footprints.

\subsection{Motion of the Ground Material}

\figureTwoCol{
  \centerline{\epsfxsize=\textwidth \epsfbox{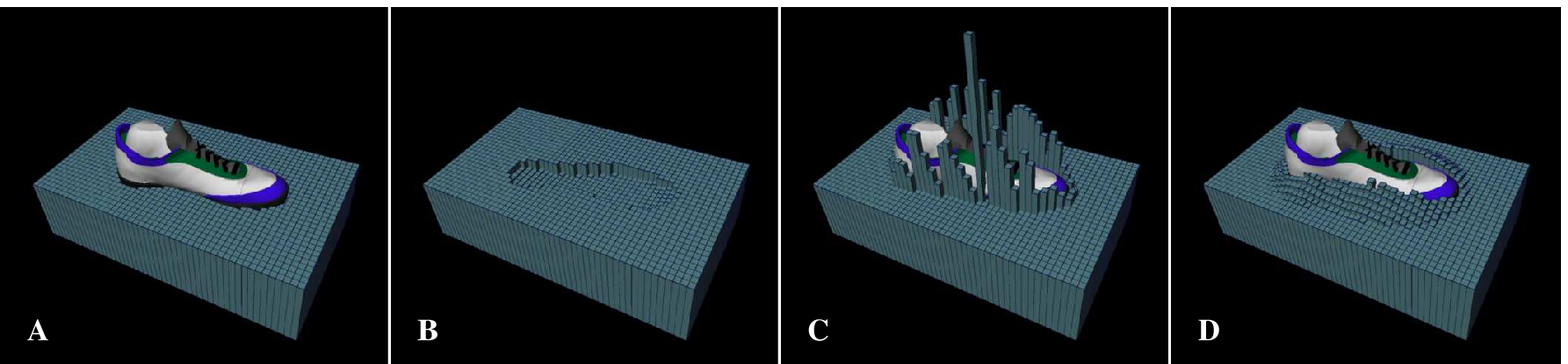}}
  \caption{
    The motion of the ground material is computed in stages.  {\em
    (A)} First, the geometric objects are intersected with the ground
    surface.  {\em (B)} Next, the penetrating columns are adjusted,
    and {\em (C)} the material is distributed to non-penetrating
    columns.  {\em (D)} Finally, the erosion process spreads the
    material.
  }\label{fig:Steps}
}

The height field represented by the top of the columns is deformed as
rigid geometric objects push into the grid.  For the examples given in
this paper, the geometric objects are a runner's shoe, a bicycle tire
and frame, and a jointed human figure.  The motion of the rigid bodies
was computed using a dynamic simulation of a human running, bicycling,
or falling down on a smooth, hard ground plane\cite{Hodgins:1995:AHA}.
The resulting motion was given as input to the simulation of the
ground material in the form of trajectories of positions and
orientations of the geometric objects.  Because of this generic
specification of the motion, the input motion need not be dynamically
simulated but could be keyframe or motion capture data.

The surface simulation approximates the motion of the columns of
ground material by compressing or displacing the material under the
rigid geometric objects.  At each time step, a test is performed to
determine whether any of the rigid objects have intersected the height
field.  The height of the affected columns is reduced until they no
longer penetrate the surface of the rigid object.  The material that
was displaced is either compressed or forced outward to surrounding
columns.  A series of erosion steps are then performed to reduce the
magnitude of the slopes between neighboring columns.  Finally,
particles can be generated from the contacting surface of the rigid
object to mimic the spray of material that is often seen following an
impact.  These stages are illustrated in figure~\ref{fig:Steps}.  We
now discuss each stage of the algorithm in more detail: collision,
displacement, erosion, and particle generation.

\paragraph* {Collision.}  The collision detection and response 
algorithm determines whether a rigid object has collided with the
ground surface.  For each column, a ray is cast from the bottom of the
column through the vertex at the top.  If the ray intersects a rigid
object before it hits the vertex, then the rigid object has penetrated
the surface and the top of the column is moved down to the
intersection point.  A flag is set to indicate that the column was
moved, and the change in height is recorded.  The computational costs
of the ray inter\-section tests are reduced by partitioning the
polygons of the rigid body models using an axis-aligned bounding box
hierarchy\cite{Snyder:1995:ITP}.

Using a vertex coloring algorithm, the simulation computes a contour
map based on the distance from each column that has collided with the
object to the closest column that has not collided
(figure~\ref{fig:GridColor}).  This information is used when the
material displaced by the collision is distributed.  As an
initialization step, columns not in contact with the object are
assigned the value zero.  During subsequent iterations, unlabeled
columns adjacent to labeled columns are assigned a value equal to the
value of the lowest numbered adjacent column plus one.

\figureSimple{
  \centerline{\epsfxsize= 0.75 \columnwidth \epsfbox{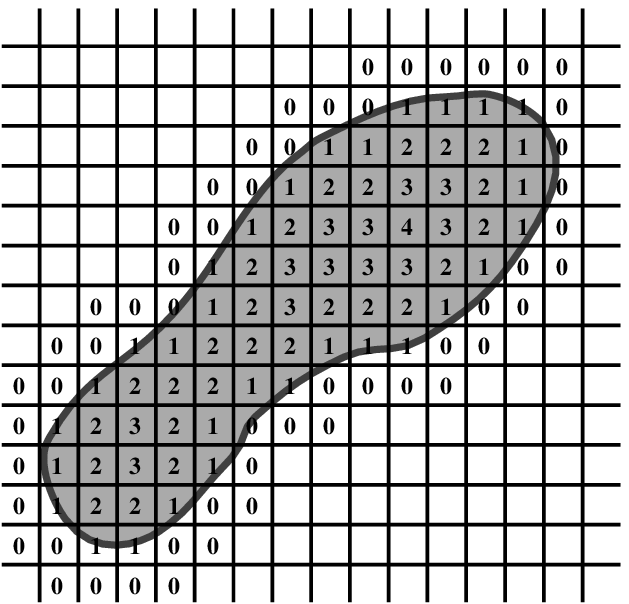}}
  \caption{
    The contour map represents the distance from each column in
    contact with the foot to a column that is not in contact.  For
    this illustration, we used columns that are four-way
    connected. However, in the examples in this paper we used
    eight-way connectivity because we found that the higher
    connectivity yielded smoother results.  
  }\label{fig:GridColor}
}

\paragraph* {Displacement.} Ground material from the
columns that are in contact with the object is either compressed or
distributed to surrounding columns that are not in contact with the
object.  The compression ratio $\alpha$ is chosen by the user and is
one of the parameters available for controlling the visual appearance
of the ground material.  The material to be distributed, $\Delta h$,
is computed by $\Delta h = \alpha m$, where $m$ is the total amount of
displaced material.  The material that is not compressed is equally
distributed among the neighbors with lower contour values, so that the
ground material is redistributed to the closest ring of columns not in
contact with the rigid object.  The heights of the columns in this
ring are increased to reflect the newly deposited material.

\paragraph* {Erosion.} Because the displacement algorithm deposits
material only in the first ring of columns not in contact with the
object, the heights of these columns may be increased in an
unrealistic fashion.  An ``erosion'' algorithm is used to identify
columns that form steep slopes with their neighbors and move material
down the slope to form a more realistic mound.  Several parameters
allow the user to control the shape of the mound and model different
ground materials.

The erosion algorithm examines the slope between each pair of adjacent
columns in the grid.  For a column $ij$ and a neighboring column $kl$,
the slope, $s$, is
\begin{equation}
  s = {\rm tan}^{-1} (h_{ij} - h_{kl}) / d
\end{equation}
where $h_{ij}$ is the height of column $ij$ and $d$ is the
distance between the two columns.  If the slope between two neighboring
columns is greater than a
threshold $\theta_{out}$, then ground material is moved from the
higher column down the slope to the lower column.  In the special case
where one of the columns is in contact with the geometric object, a
different threshold, $\theta_{in}$, is used to provide independent
control of the inner slope.  
Ground material is
moved by computing the average difference in height, $\Delta h_a$, for
the $n$ neighboring columns with too large a downhill slope:
\begin{equation}
  \Delta h_a = {{\sum (h_{ij} - h_{kl})} \over {n}}.
\end{equation}
The average difference in height is multiplied by a fractional
constant, $\sigma$, and the resulting quantity is equally distributed
among the downhill neighbors.  
This step in the algorithm repeats until all slopes
are below a threshold, $\theta_{stop}$.  
The erosion algorithm may cause some
columns to intersect a rigid object but this penetration will be
corrected on the next time step.

\figureTwoCol{
  \vskip 2ex
  \centerline{%
    \hfill
    {\epsfysize= 2in \epsfbox{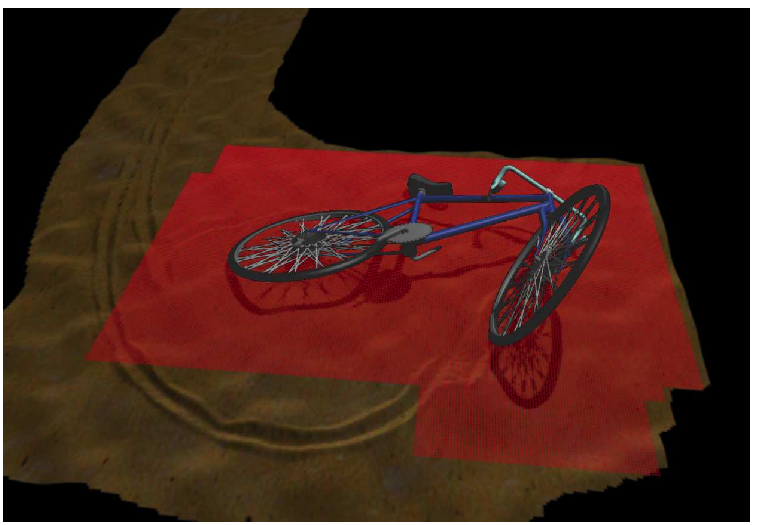}}
    \hfill
    {\epsfysize= 2in \epsfbox{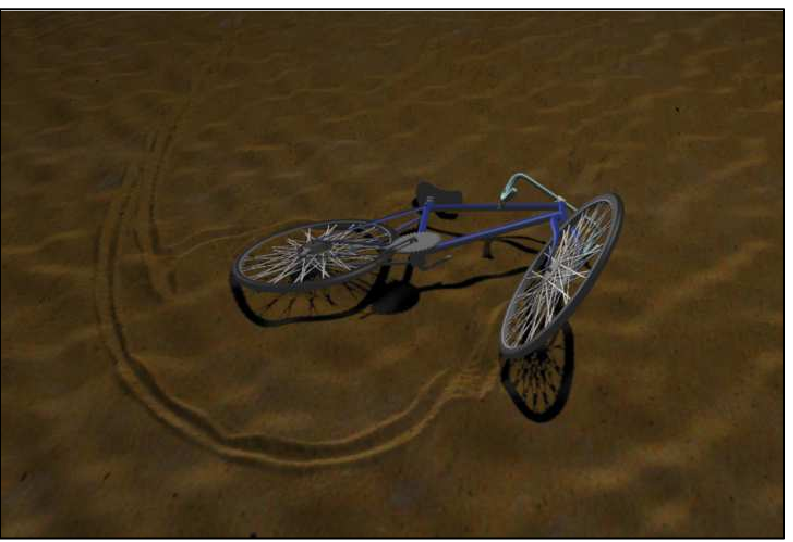}}
    \hfill
  } 
  \caption{
    The left figure shows the ground area that has been created in the
    hash table.  The currently active area is highlighted in red.  The
    right figure shows the same scene rendered over the initial ground
    surface.  There are approximately {37,000} columns in the active
    area and {90,000} stored in the hash table; the number of
    columns in the entire virtual grid is greater than $2$~million.
  }\label{fig:Bike}
}

\paragraph*{Particle Generation.}
We use a particle system to model portions of the ground material that
are thrown into the air by the motion of the rigid geometric objects.
The user controls the adhesiveness between the object and the material
as well as the rate at which the particles fall from the object.  Each
triangle of the object that is in contact with the ground picks up a
volume of the ground material during contact.  The volume of material
is determined by the area of the triangle multiplied by an adhesion
constant for the material.  When the triangle is no longer in contact
with the ground, it drops the attached material as particles according
to an exponential decay rate.
\begin{equation}
  \Delta v = v ( e^{(-t + t_c + \Delta t) / h } - e^{(-t + t_c) / h} )
\end{equation}
where $v$ is the initial volume attached to the triangle, $t$ is the
current time, $t_c$ is the time at which the triangle left the ground,
$\Delta t$ is the time step size, and $h$ is a half life parameter
that controls how quickly the material falls off.  The number of
particles released on a given time step is determined by 
$n = \Delta v / \phi$, where ${\phi}$ is the volume of each particle.

The initial position, ${\bf p_0}$, for a particle is randomly
distributed over the surface of the triangle according to:
\begin{equation}
	{\bf p_0} = b_a {\bf x}_a  + b_b {\bf x}_b  + b_c {\bf x}_c 
\end{equation}
where ${\bf x}_a$, ${\bf x}_b$, and ${\bf x}_c$ are the coordinates of
the vertices of the triangle and $b_a$, $b_b$, and $b_c$ are the
barycentric coordinates of ${\bf p_0}$ given by
\begin{eqnarray}
	b_a &=& 1.0 - \sqrt{ \rho_a } \\
	b_b &=& \rho_b ( 1.0 - b_a ) \\
	b_c &=& 1.0 - ( b_a + b_b )
\end{eqnarray}
where $\rho_a$ and $\rho_b$ are independent random variables evenly
distributed between $[0..1]$.  
This computation results in a uniform
distribution over the triangle\cite{Turk:1990:GRP}.

The initial velocity of a particle is computed from the velocity of
the rigid object:
\begin{equation}
	{\bf \dot{p}}_0 = \mbox{\boldmath $\nu$} + \mbox{\boldmath $\omega$} \times {\bf p}_0
\end{equation}
where \mbox{\boldmath $\nu$} and \mbox{\boldmath $\omega$} are the
linear and angular velocity of the object.  To give a more realistic
and appealing look to the particle motion, the initial velocities are
randomly perturbed.

The final component of the particle creation algorithm accounts for
the greater probability that material will fall off rapidly
accelerating objects.  A particle is only created if
$(|{\bf\ddot{p}}_0|/s)^\gamma > \rho$, where $s$ is the minimal
acceleration at which all potential particles will be dropped,
$\gamma$ controls the variation of the probability of particle
creation with speed, and $\rho$ is a random variable evenly
distributed in the range $[0..1]$.

If particles are only generated at the beginning of a time step then
the resulting particle distribution will have a discrete, sheetlike
appearance.  We avoid this undesirable effect by randomly distributing
each particle's creation time within the time step interval.  The
information used to calculate the initial position and velocity is
interpolated within the interval to obtain information appropriate for
the particle's creation time.

Once generated, the particles fall under the influence of gravity.
When a particle hits the surface of a column, its volume is added to
the column.


\subsection {Implementation and Optimization} Simulations of terrain
generally span a large area.  For example, we would like to be able to
simulate a runner jogging on a beach, a skier gliding down a
snow-covered slope, and a stampede of animals crossing a sandy valley.
A naive implementation of the entire terrain would be intractable
because of the memory and computation requirements.  The next two
sections describe optimizations that allow us to achieve reasonable
performance by storing and simulating only the active portions of the
surface and by parallelizing the computation.

\paragraph*{Algorithm Complexity.} Because the ground model is a
two-dimensional rectilinear grid, the most straightforward
implementation is a two-dimensional array of nodes containing the
height and other information about the column.  If an animation
required a grid of \(i\) rows and \(j\) columns, \(i \times j \) nodes
would be needed, and computation time and memory would grow linearly
with the number of grid points.  Thus, a patch of ground 10\,m $\times$
10\,m with a grid resolution of 1\,cm yields a \(1000 \times 1000
\) grid with one million nodes.  If each node requires 10\,bytes of
memory, the entire grid requires 10\,Mbytes of storage.  Even this
relatively small patch of ground requires significant system
resources.  However, most of the ground nodes are static throughout
the simulation, allowing us to use a much more efficient algorithm
that creates and simulates only the active nodes.

The active area of the ground surface is determined by projecting an
enlarged bounding box for the rigid objects onto the surface as shown
in figure~\ref{fig:Bike}.  The nodes within the projection are marked
as active, and the collision detection, displacement, and erosion
algorithms are applied, not to the entire grid, but only to these
active grid points.  Additionally, nodes are not allocated for the
entire ground surface, rather they are created on demand as they become
active.  The $ij$ position of a particular node is used as the
index into a hash table allowing the algorithms to be implemented as
if a simple array of nodes were being used.

Because only the active grid points are processed, the computation
time is now a function of the size of the rigid objects in the scene
rather than the total grid size.  Memory requirements are also
significantly reduced, although the state of all modified nodes must
be stored even after they are no longer active.

\figureSimple{
  \vskip -.15in
  \centerline{\epsfxsize=\columnwidth \epsfbox{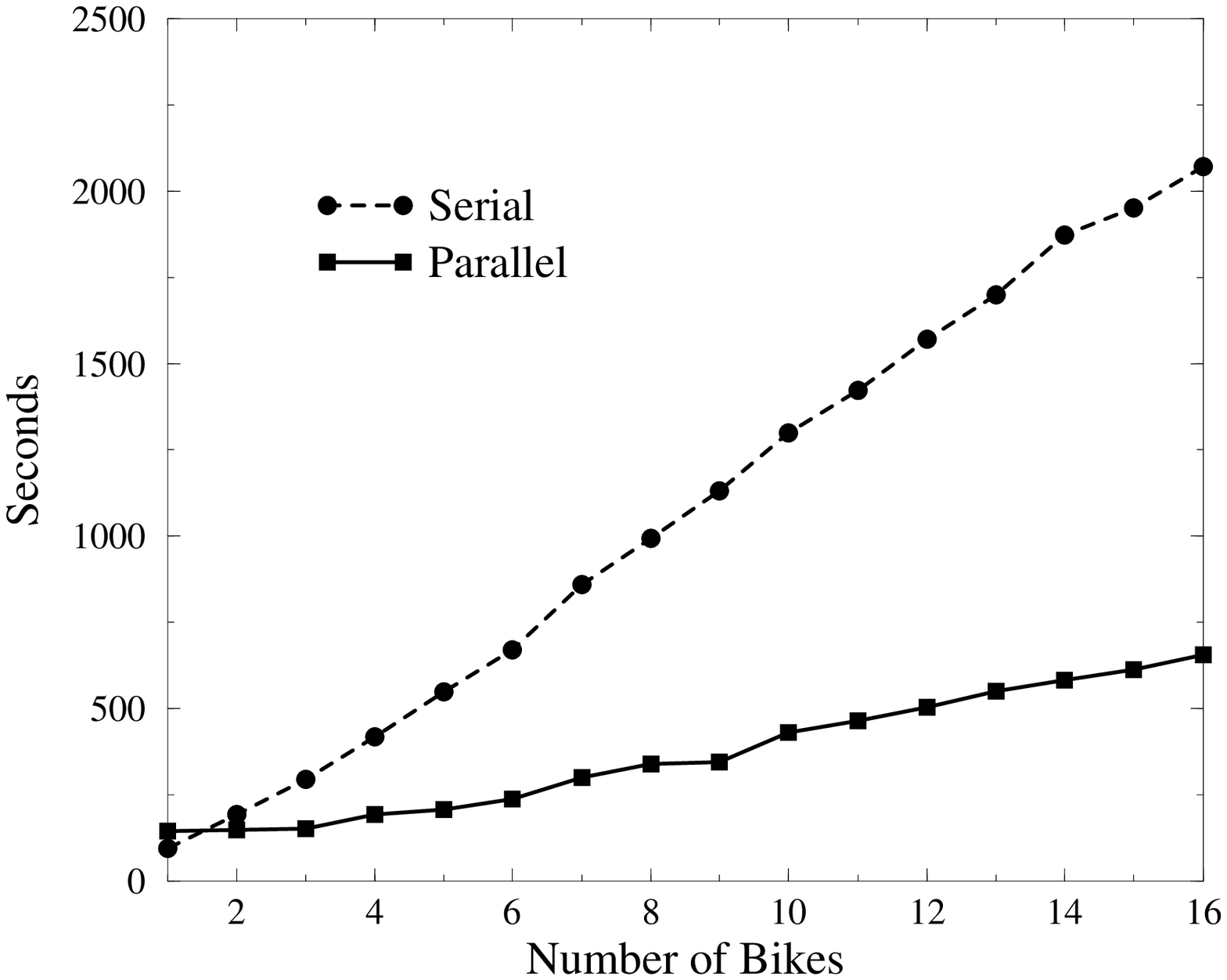}}
  \vskip -.25in
  \caption{
    These timing results were computed on a Silicon Graphics Power
    Challenge system with $16$ $195$MHz MIPS R10000 processors and
    4\,Gbytes of memory.  Each character is an alien biker like the ones
    shown in figure~\ref{fig:AlienBikes}.  Times plotted are for one
    second of simulated motion.
  }\label{fig:Graph}
}

\paragraph*{Parallel Implementation.} Despite the optimization provided
by simulating only active nodes, the computation time grows linearly
with the projected area of the rigid objects.  Adding a second
character will approximately double the active area (see
figure~\ref{fig:Graph}), but the computation time for multiple
characters can be reduced by using parallel processing when the
characters are contacting independent patches of ground.

We have designed and implemented a parallel scheme for the ground
surface simulation.  A single parent process maintains the state of
the grid and coordinates the actions of the child processes.  During
initialization, a child process is created for each character that
will interact with the ground surface.  The children communicate with
the parent process via the UNIX socket mechanism and may exist
together on a single multiprocessor machine or on several separate
single processor machines.

Each child computes the changes to the grid caused by its character as
quickly as possible, without any direct knowledge about the progress
of the other children.  When a child completes the computation for a
time step, it reports the changes it has made to the parent process
and then waits for information about any new grid cells that will be
in the bounding box for its character during the next time step.
However, if the child is ready to compute a time step before another
child has reported prior changes that are within the bounding box of a
character assigned to the first child, the parent will prevent the
first child from continuing until the changes are available.  For
example, in an animation of a cyclist that rides across a footprint
left by a runner, the child process computing the cyclist may arrive
at the footprint location before the process computing the runner has
simulated the creation of the footprint.  If two or more characters
have overlapping bounding boxes for the same time step, the
computation for those characters is reassigned to a single child
process until they no longer overlap.

\figureTwoCol{
  \vskip 0.5ex
  \centerline{\epsfxsize=6in \epsfbox{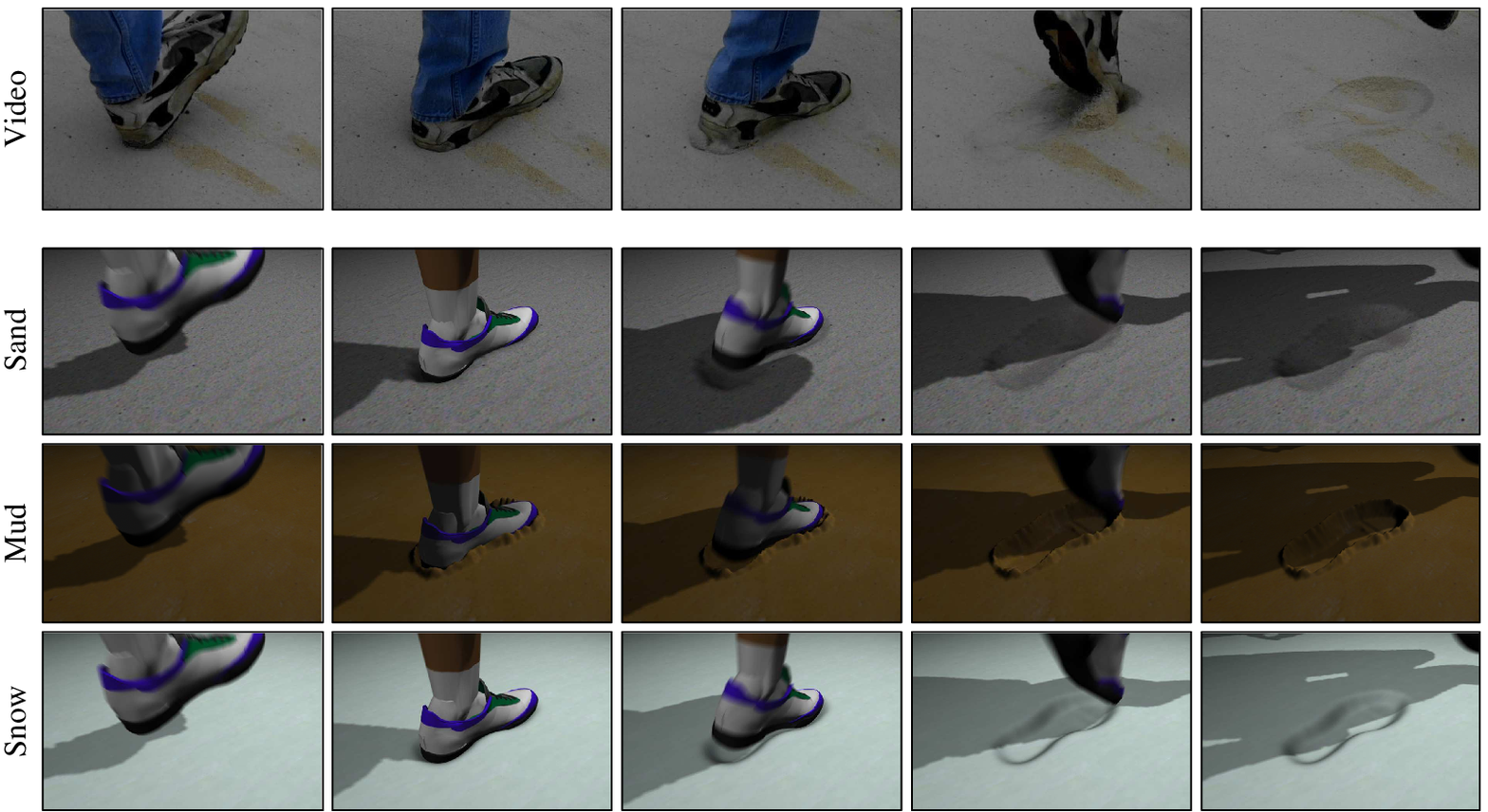}}
  \vskip -2ex
  \caption{
    Images from video footage of a human runner stepping in sand and a
    simulated runner stepping in sand, mud, and snow.  The human
    runner images are separated by 0.133\,s; the simulated images are
    separated by 0.1\,s.  
  }\label{fig:RsmsFilmStrip}
}

\figureTwoCol{
  \vskip 0.5ex
  \centerline{\epsfxsize=6in \epsfbox{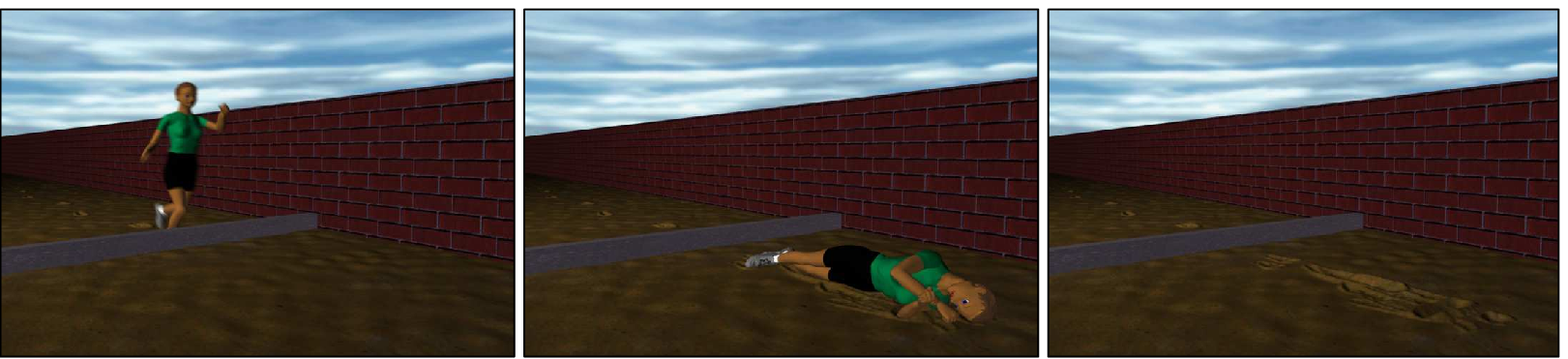}}
  \vskip -2ex
  \caption{
    Images of a runner tripping over an obstacle and falling onto the
    sand.  The final image shows the pattern she made in the
    sand.
  }\label{fig:Trip}
}

Simulation run times for both the serial and parallel versions of our
algorithm are shown in figure~\ref{fig:Graph}.  As expected, the time
required by the serial implementation grows linearly with the number
of characters.  Ideally, the time required by the parallel
implementation would be constant since each character has its own
processor.  However, due to communication overhead and interactions
between the characters, the run time for the parallel
version grows as the number of characters increases, but at a much
slower rate than the serial version.


\section {Animation Parameters} 

One goal of this research is to create a tool that allows animators to
easily generate a significant fraction of the variety seen in ground
materials.  Five parameters of the simulation can be changed by the
user to achieve different effects: inside slope, outside slope, 
roughness, liquidity, and compression.  The first four are used by the
erosion algorithm, and the fifth is used by the displacement
algorithm.

The inside and outside slope parameters, $\theta_{in}$ and
$\theta_{out}$, modify the shape of a mound of ground material by
changing the slope adjacent to intersecting geometry and the slope on
the outer part of the mound.  Small values lead to more erosion and a
more gradual slope; large values yield less erosion and a steeper
slope.

Roughness, $\sigma$, controls the irregularity of the ground
deformations by changing the amount of material that is moved from one
column to another during erosion.  Small values yield a smooth mound
of material while larger values give a rough, irregular surface.

Liquidity, $\theta_{stop}$, determines how watery the material appears
by controlling the amount of erosion within a single timestep.  With
less erosion per time step, the surface appears to flow outward from
the intersecting object; with more erosion, the surface moves to its
final state more quickly.

The compression parameter, $\alpha$, offers a way to model substances
of different densities by determining how much displaced material is
distributed outward from an object that has intersected the grid.  A
value of one causes all material to be displaced; a value less than
one allows some of the material to be compressed.

When particles are used, additional parameters are required to
determine their appearance.  We included parameters to control
adhesion, particle size, and the rate at which material falls off of
the objects.  We used particles in the animations of sand but did not
include them in the animations of mud or snow.  Other more dynamic
motions such as skiing might generate significant spray, but running
in snow appears to generate clumps of snow rather than particles.


\section {Results and Discussion}

\tableSimple{
  \begin{center}
    \begin{tabular} {|l|l|l|l|l|} \hline
      \multicolumn{1}{|l|}{Effect} & \multicolumn{1}{l|}{Variable} & \multicolumn{1}{l|}{Sand} & \multicolumn{1}{l|}{Mud} & \multicolumn{1}{l|}{Snow} \\ \hline
      inside slope 	& $\theta_{in}$ 	&  0.8   &      1.57   &     1.57 \\
      outside slope 	& $\theta_{out}$ 	&  0.436 &      1.1    &     1.57 \\
      roughness  	& $\sigma$     		&  0.2   &      0.2    &     0.2  \\ 
      liquidity 	& $\theta_{stop}$	&  0.8   &      1.1    &     1.57 \\
      compression 	& $\alpha$   		&  0.3   &      0.41   &     0.0  \\ \hline
    \end{tabular}
  \end{center}
  \caption{
    Table of parameters for the three ground
    materials.  
  }\label{tab:ParameterTable}
}

Figure~\ref{fig:RsmsFilmStrip} shows images of a human runner stepping in
sand and a simulated runner stepping in sand, mud, and snow.  The
parameters used for the simulations of the three ground materials are
given in table~\ref{tab:ParameterTable}.  The footprints left by the
real and simulated runners in sand are quite similar.

Figures~\ref{fig:Bike} and~\ref{fig:Trip} show more complicated
patterns created in the sand by a falling bicycle and a tripping
runner.  For each of these simulations, we used a grid resolution of
1\,cm by 1\,cm yielding a virtual grid size of 2048$\times$1024 for the
bicycle and 4096$\times$512 for the runner.

The images in this paper were rendered with Pixar's RenderMan
software.  We found that rendering the ground surface using a
polygonal mesh was computationally expensive and that the data files
required to describe the mesh were large and difficult to work with.
We achieved better results using a single polygon with a
displacement shader that modeled the ground surface.

The simulation described in this paper allows us to capture many of
the behaviors of substances such as sand, mud, and snow.  Only about
fifteen iterations were required to hand tune the parameters for the
desired effect with each material.  The computation time is not
burdensome: a 3-second simulation of the running figure interacting
with a 1\,cm by 1\,cm resolution ground material required less than
2\,minutes of computation time on a single $195$MHz MIPS R10000
processor.

Many effects are missed by this model.  For example, wet sand and
crusty mud often crack and form large clumps, but our model can
generate only smooth surfaces and particles.  Actual ground material
is not uniform but contains both small grains of sand or dirt as well
as larger objects such as rocks, leaves, and sea shells.  More
generally, many factors go into creating the appearance of a given
patch of ground: water and wind erosion, plant growth, and the
footprints of many people and animals.  Some of these more subtle
effects are illustrated by the human footprints in snow and mud shown
in figure~\ref{fig:Reality}.

\figureSimple{
  \centerline{\epsfxsize=3.in \epsfbox{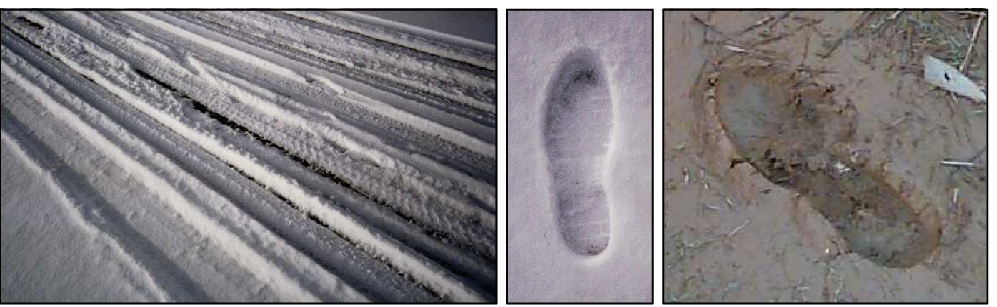}}
  \caption{
    Images of actual tire tracks in snow and human footprints in snow
    and in mud.  
  }\label{fig:Reality}
}

One significant approximation in the ground simulation is that the
motion of the rigid objects is not affected by the deformations of the
surface.  For the sequences presented here, each of the rigid body
simulations interacted with a flat, smooth ground plane.  A more
accurate and realistic simulation would allow the bike and runner to
experience the undulations in the initial terrain as well as the
changes in friction caused by the deforming surfaces.  For example, a
bike is slowed down significantly when rolling on sand and a runner's
foot slips slightly with each step on soft ground.

Other approximations are present in the way that the sand responds to
the motion of the rigid objects.  For example, a given area of sand has
no memory of the compression caused by previous impacts.  Because the
motion of the rigid objects are specified in advance, this
approximation does not cause any noticeable artifacts.  Compression
could also be used to change rendering parameters as appropriate.  

We do not take the velocity of the rigid object into account in the
ground simulation.  For bicycling and running, this approximation is
negligible because the velocities of the wheel and the foot with respect
to the ground are small at impact.  For the falling runner or bicycle,
however, this approximation means that the ridge of sand is uniformly
distributed rather than forming a larger ridge in the direction of
travel.

The motions of sand, mud, and snow that we generated are distinctly
different from each other because of changes to the simulation
parameters.  Although much of the difference is due to the
deformations determined by our simulations, part of the visual
difference results from different surface properties used for
rendering.  To generate the images in this paper, we had not only to
select appropriate parameters for the simulation but also to select
parameters for rendering.  A more complete investigation of techniques
for selecting rendering parameters and texture maps might prove
useful.

We regard this simulation as appearance-based rather than
engineering-based because most of the parameters bear only a scant
resemblance to the physical parameters of the material being modeled.
The liquidity parameter, for example, varies between $0$ and $\pi/2$
rather than representing the quantity of water in a given amount of
sand.  It is our hope that this representation for the parameters
allows for intuitive adjustment of the resulting animation without
requiring a deep understanding of the simulation algorithms or soil
mechanics.  The evaluation is also qualitative or appearance-based in
that we compare simulated and video images of the footprints rather
than matching initial and final conditions quantitatively.


\section*{Acknowledgments} 

A previous version of this paper appeared in \textit{The
Proceedings of Graphics Interface '98}.

This project was supported in part by NSF NYI Grant No. IRI-9457621
and associated REU funding,
Mitsubishi Electric Research Laboratory, and a Packard Fellowship. The
second author was supported by an
Intel Fellowship.



\bibliographystyle{cgf}
\bibliography{sand}


\vfill\eject 
\end{document}